\documentstyle[12pt,a4wide]{article}
\input epsf
\newcommand{\be}{\begin{equation}}
\newcommand{\ee}{\end{equation}}

\def\ben{\begin{equation}}
\def\een{\end{equation}}
\def\bea{\begin{eqnarray}}
\def\eea{\end{eqnarray}}
\begin{document}

\title{\vskip -60pt\begin{flushright}
{\normalsize To appear in Phys. Lett. B.} \\
\end{flushright}\vskip 50pt
\bf \large \bf SKYRMED MONOPOLES\\[30pt]
\author{D.{{Yu.}} Grigoriev$^{1}$\thanks{ On leave of absence from
Institute for Nuclear Research of Russian Academy of Sciences}\ ,
 P.M. Sutcliffe$^{2}$ and D.H. Tchrakian$^{1}$\\[10pt]
\\{\normalsize $^{1}$ {\sl  Mathematical Physics,
National University of Ireland Maynooth (NUIM),   }}
\\{\normalsize {\sl  Maynooth, Co. Kildare, Ireland.}}
\\{\normalsize {\sl Email : dima@thphys.may.ie, 
 tigran@maxwell.thphys.may.ie  }}\\
\\{\normalsize $^{2}$  {\sl Institute of Mathematics,
 University of Kent at Canterbury,}}\\
{\normalsize {\sl Canterbury, CT2 7NF, U.K.}}\\
{\normalsize{\sl Email : P.M.Sutcliffe@ukc.ac.uk}}\\}}
\date{June 2002}
\maketitle
 
\begin{abstract}
We investigate multi-monopole solutions of a modified version of
the BPS Yang-Mills-Higgs model in which a term quartic in the
covariant derivatives of the Higgs field (a Skyrme term) is included in
the Lagrangian. Using numerical methods we find that this modification
leads to multi-monopole bound states. We compute axially symmetric
monopoles  up to charge five and also monopoles with Platonic symmetry for
charges three, four and five. The numerical evidence suggests that, in
contrast to Skyrmions, the minimal energy Skyrmed monopoles are axially
symmetric.

\end{abstract}
\newpage
 
\section{Introduction}
Two of the most interesting kinds of topological solitons in three
space dimensions are BPS monopoles and Skyrmions. Although there are
some similarities between monopoles and Skyrmions, which we shall
discuss shortly, there are a number of important differences which
we first recall. $SU(2)$ BPS monopoles are soliton solutions of a
 Yang-Mills-Higgs gauge theory (with a massless Higgs) in which the 
topological charge $N$ is an element of the second homotopy
group of the two-sphere, identified as the Higgs field vacuum manifold.
The topological charge is therefore associated with a winding of the
Higgs field on the two-sphere at spatial infinity. In the BPS limit
there is a $4N$-dimensional moduli space of static solutions which
are degenerate in energy, so in this sense there are no stable bound
states since any static charge $N$ solution has the same energy as
$N$ well-separated charge one monopoles. Contrast these features with
those of Skyrmions, which are soliton solutions of a nonlinear sigma
model with target space $SU(2).$ The Skyrme field is constant on the
 two-sphere at spatial infinity and this yields a compactification
of Euclidean three-space to a three-sphere. The integer-valued topological
charge (baryon number) is an element of the third homotopy group of
the target space and counts the number of times that the target space
is covered by the Skyrme field throughout space. There are static forces
between Skyrmions, which for a suitable relative internal orientation
are attractive, and this leads to multi-Skyrmion bound states.

To summarize, three main differences between Skyrmions and monopoles
are the basic fields of the model, the way the topological charge arises,
and the existence (or not) of bound states. Given these facts it is
rather surprising that there appears to be some similarity between
various monopole and Skyrmion solutions. There are axially symmetric
monopoles and Skyrmions for all charges greater than one (although above
charge two these are not the minimal energy Skyrmions) and both have
solutions with Platonic symmetries for the same certain charges.
For example, there is a tetrahedral monopole for $N=3,$ a cubic
monopole for $N=4$ and a dodecahedral monopole for $N=7$ \cite{HMM,HS}.
All { BPS} monopoles of a given charge have the same energy but these 
particular monopole solutions are selected out by being mathematically
more tractable than an arbitrary solution. For these three values of the
charge $N=3,4,7$ 
the minimal energy Skyrmion has precisely the same symmetry
as the above monopoles and energy density isosurfaces are
qualitatively similar \cite{BTC,BS}. These and other similarities
can be partially understood by relating both types of soliton to
rational maps between Riemann spheres \cite{HMS}. 

\noindent The obvious differences and yet remarkable similarities between
monopoles and Skyrmions is the motivation for the present work,
where we aim to modify the BPS monopole Lagrangian by the
addition of a Skyrme-like term
with the goal of breaking the energy degeneracy and producing
monopole bound states. That such a modification might yield 
 monopole bound states is suggested by the fact that more complicated
models, involving Skyrme-like terms, have been shown to have 
this property \cite{KOT}.

We refer to the soliton solutions of our modified model as
Skyrmed monopoles, though in the following for brevity we mainly use the
term monopoles, and refer to monopole solutions of the unmodified model
as BPS monopoles. Our numerical computations, for monopoles
up to charge five, show that the modified model does indeed have
multi-monopole bound states, though perhaps surprisingly our numerical
results suggest that the minimal energy multi-monopoles are all
axially symmetric and do not share the Platonic symmetries of
the corresponding minimal energy Skyrmions. Platonic monopole solutions
are computed, and although they have low energies they are very slightly 
above those of the axially symmetric solutions.

Explicitly, the model we consider is defined by the following
energy function (we deal only with static solutions in this letter
but the extension to the relativistic Lagrangian is obvious)
\be
E=\frac{1}{8\pi}\int -\mbox{Tr}\bigg(
\frac{1}{2}F_{ij}F_{ij}+ D_i\Phi D_i\Phi
+\frac{\mu^2}{2}[D_i\Phi, D_j\Phi][D_i\Phi, D_j\Phi]\bigg)
\, d^3x.
\label{energy}
\ee
Here Latin indices run over the spatial values 1,2,3, the
Higgs field and gauge potential are $\Phi, A_i\in su(2),$
the covariant derivative is $D_i\Phi=\partial_i\Phi+[A_i,\Phi]$ 
and $F_{ij}$ is the field strength. 

The boundary condition is that  $|\Phi|^2=-\frac{1}{2}\mbox{Tr}\Phi^2$
equals one at spatial infinity. The Higgs field at infinity then defines
a map between two-spheres and the winding number of this map is the 
monopole number $N.$

If $\mu=0$ then the energy (\ref{energy}) is the usual BPS
Yang-Mills-Higgs energy and monopole solutions satisfy the first
order Bogomolny equations. All members of the $4N$-dimensional moduli
space of solutions have energy $E=N,$ and include solutions describing
$N$ well-separated monopoles as well as axially symmetric $N$-monopoles. 
For $\mu\ne 0$ the additional term is the gauge analogue of the Skyrme
term for the sigma model. In the sigma model context the presence of
the Skyrme term is necessary to have stable
soliton solutions but in the monopole
context it is optional. Clearly the energy degeneracy of the BPS model
will be broken for $\mu\ne 0$ and as we shall describe below this
produces monopole bound states, rather than the familiar
monopole-monopole repulsion induced by the addition of a Higgs potential,
which is an alternative  way to lift the energy degeneracy of the 
BPS model.

\section{Numerical Methods and Results}

In order to construct static solutions of the field equations which
follow from the variation of the energy (\ref{energy}) we apply a
simulated annealing algorithm \cite{sa} to minimize the energy using a
finite difference discretization on a grid containing $81^3$ points
with a lattice spacing $dx=0.25.$ Note that this grid is a little
smaller than those currently in use to study similar problems for
Skyrmions \cite{BS} (although simulated annealing computations
of Skyrmions on grids containing $80^3$ points do provide accurate
results \cite{Hale}), but the
large number of fields which need to be dealt with in studying a
Yang-Mills-Higgs gauge theory make it difficult to handle grids much
larger than this with our current resources. However, by testing our
codes on the BPS limit ($\mu=0$) where exact results are known, we are
able to estimate the numerical errors involved and have confidence in
our results being accurate to the level that we discuss later.

In order to apply our annealing code we need to provide initial conditions
which have the correct topological winding of the Higgs field at infinity.
To provide these initial conditions, and be able to prescribe any 
particular symmetry that we may want to impose, we make use of a formula
relating the asymptotic Higgs field to a rational map between Riemann
spheres \cite{IS1}. 
Explicitly, the initial Higgs field is given by
\be
\Phi=
\frac{if(r)}{1+\vert R\vert^2}
\pmatrix{1-\vert R\vert^2& 2\bar R\cr
2R & \vert R\vert^2-1\cr}
\label{higgsic}
\ee where $f(r)$ is a real profile function, which depends on the
radius $r,$ and satisfies the boundary conditions that $f(0)=0$ and
$f=1$ on the boundary of the numerical grid. Here $R(z)$ is a rational
map of degree $N$ in the complex variable $z,$ ie. a ratio of two
polynomials of degree no greater than $N,$ which have no common
factors and at least one of the polynomials has degree precisely $N.$
The variable $z$ is a Riemann sphere coordinate on the unit sphere
around the origin in space ie. it is given by
$z=e^{i\phi}\tan(\theta/2)$ where $\theta$ and $\phi$ are the usual
polar coordinates.  In the BPS case there is a one-to-one
correspondence between charge $N$ monopole solutions and (an
equivalence class of) degree $N$ rational maps \cite{Ja} and the
existence of certain symmetric monopole solutions can be proved by the
construction of the associated symmetric maps \cite{HMS}. Although
there is clearly no such correspondence in our modified model we shall
make use of some of the relevant symmetric maps in our initial
conditions. 
We take all gauge potentials to
be zero initially and this preserves any symmetry that the Higgs field
may initially have. On the boundary of the grid the Higgs field is
fixed to the initial form (\ref{higgsic}), which in particular ensures
that the winding number remains equal to $N,$ but the gauge potential
is annealed to minimize the energy given the fixed boundary Higgs
field.

As a test of the accuracy of our code we first compute several BPS 
monopoles. For the $N=1$ monopole (with rational map $R=z$) 
we find an energy $E=1.007$ whose
deviation from unity is an indication of the error associated with 
the energy values we quote. Another important test is to compare 
the energies of different BPS multi-monopole solutions which
have the same charge. Of course a perfect calculation would produce
energies equal to the charge for any solution. As an example, using
the rational map $R=z^3$ of the axially symmetric 3-monopole in
the initial condition produces the energy $E=3.021,$ whereas
the rational map 
$R=(\sqrt{3}iz^2-1)/(z^3-\sqrt{3}iz)$
anneals to produce a tetrahedrally symmetric 3-monopole with energy
$E=3.018.$
This illustrates the fact that our energies are accurate to around
$1\%$ but that comparisons between different configurations are likely
to be more accurate, in this case the error is around $0.1\%.$
Similar results were obtained for other BPS examples.
\begin{figure}[ht]
\begin{center}
\leavevmode
\epsfxsize=15cm\epsffile{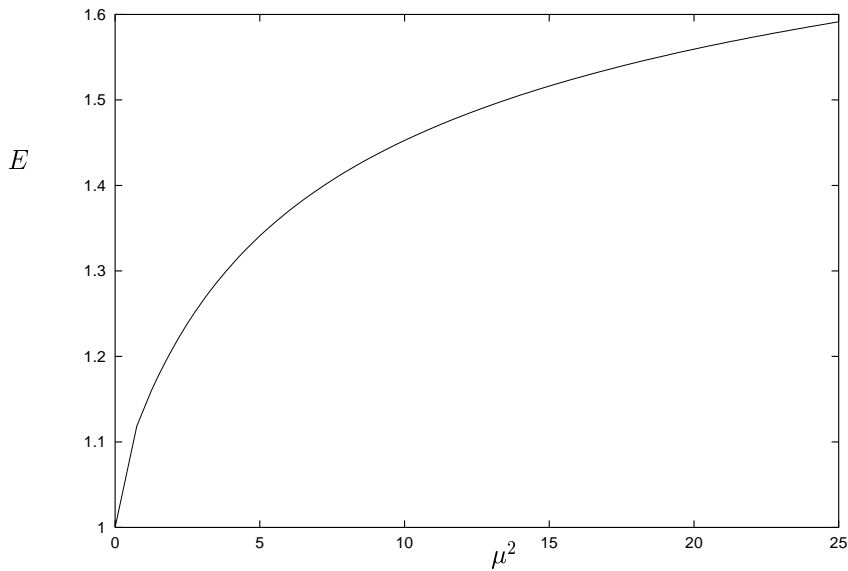}
\ \vskip -10cm
\caption{The 1-monopole energy as a function of $\mu^2$.}
\end{center}
\end{figure}

We now turn to the modified model with $\mu\ne 0,$ and the first 
issue to address is a suitable choice for the value of $\mu.$
To facilitate numerical comparisons it is useful to choose
a value of $\mu$ large enough so that the additional term raises
the energy of the $N=1$ monopole 
by something of the order of $50\%$ from the BPS value,
since it then has an effect significant enough to be calculated
numerically but does not dominate over the usual terms.
In fig.~1 we plot the energy of the $N=1$ monopole as a function
of $\mu^2.$ This calculation is performed by using a hedgehog ansatz
and computing the energy minimizing profile functions.
From fig.~1 we see that a reasonable choice is $\mu=5,$ which
we use from now on, and this gives $E=1.591.$
Using the full three-dimensional annealing code we compute the 1-monopole
energy to be $E_1=1.602$ which is in reasonable agreement with the more
accurate one-dimensional calculation.

The crucial calculation is now to compute the energy of the axially
symmetric 2-monopole. Using the rational map $R=z^2$ we compute the
axially symmetric 2-monopole, whose energy density isosurface is
displayed in fig.~2A, and find the energy $E_2=2.777.$ The important
point is that $E_2/2=1.388<1.602=E_1$ so a 2-monopole bound state
exists. It seems reasonable to conclude that the minimal energy
2-monopole is axially symmetric, though clearly we have not proved this.
 Note that $2E_1-E_2=0.427$ and, as we mentioned above,
this is expected to be significantly larger than the numerical errors
present in our energy comparisons. If required a more accurate
calculation of the 2-monopole energy could be performed by making use
of the axial symmetry to reduce to an effective two-dimensional
computation.

\begin{table}
\centering
\begin{tabular}{|c|c|c|c|}
\hline
$N$ & $G$ & $E$  & $E/N$\\
\hline 
1 & $O(3)$ & 1.602 & 1.602\\
2 & $O(2)\times Z_2$ & 2.777 & 1.388\\
3 & $O(2)\times Z_2$ & 3.807 & 1.269\\
3 & $T_d$ & 3.869 & 1.290\\
4 & $O(2)\times Z_2$ & 4.847 & 1.212\\
4 & $O_h$ & 4.974 & 1.244\\
5 & $O(2)\times Z_2$ & 5.924 & 1.185\\
5 & $D_{2d}$ & 5.982 & 1.196\\
5  & $O_h$ & 5.987 & 1.197\\
\hline
\end{tabular}\label{table}
\caption{The monopole charge $N,$ the symmetry group $G$ of the
energy density, the energy $E$ and energy per monopole $E/N$ for
several examples of Skyrmed monopoles.}
\end{table}

For higher charges we first look at axially symmetric monopoles by
using the rational maps $R=z^N.$ For $N=2,3,4,5$ the energies $E_N$ 
and energies per monopole $E_N/N$ are presented in Table 1
and we display energy density isosurfaces in figs.~2A,2B,2C,2D.
 We also plot the
energy per monopole for these axially symmetric solutions as a function
of monopole number in fig.~3. This plot demonstrates that all these
solutions are stable against the break-up into $N$ well separated
monopoles, and also into any well-separated clusters containing 
single or axially symmetric monopoles. Note that for these axially symmetric
solitons the energy per monopole
{\em decreases} as the monopole number increases and this contrasts
sharply with Skyrmions. For axially symmetric Skyrmions with $N\ge 2$
the energy per Skyrmion  {\em increases} with the number of Skyrmions
\cite{KS}, and only the $N=2$ minimal energy Skyrmion has an 
axial symmetry. Furthermore, for $N>4$ the axially symmetric
charge $N$ Skyrmion is not even bound against the break-up into
$N$ well-separated single Skyrmions.

\begin{figure}[ht]
\begin{center}
\leavevmode
\epsfxsize=15cm\epsffile{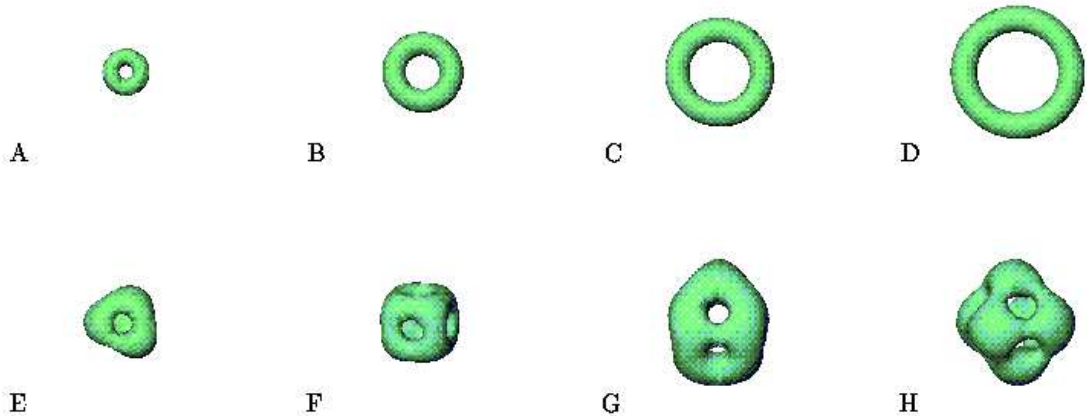}
\caption{Energy density isosurfaces (to scale) of various Skyrmed monopoles.
\quad \quad \quad \quad\quad \quad \quad \quad
A) $N=2$ axial,
B) $N=3$ axial,
C) $N=4$ axial,
D) $N=5$ axial,
E) $N=3$ tetrahedral,
F) $N=4$ octahedral,
G) $N=5$ dihedral,
H) $N=5$ octahedral.}
\end{center}
\end{figure}

\begin{figure}[ht]
\begin{center}
\leavevmode
\epsfxsize=15cm\epsffile{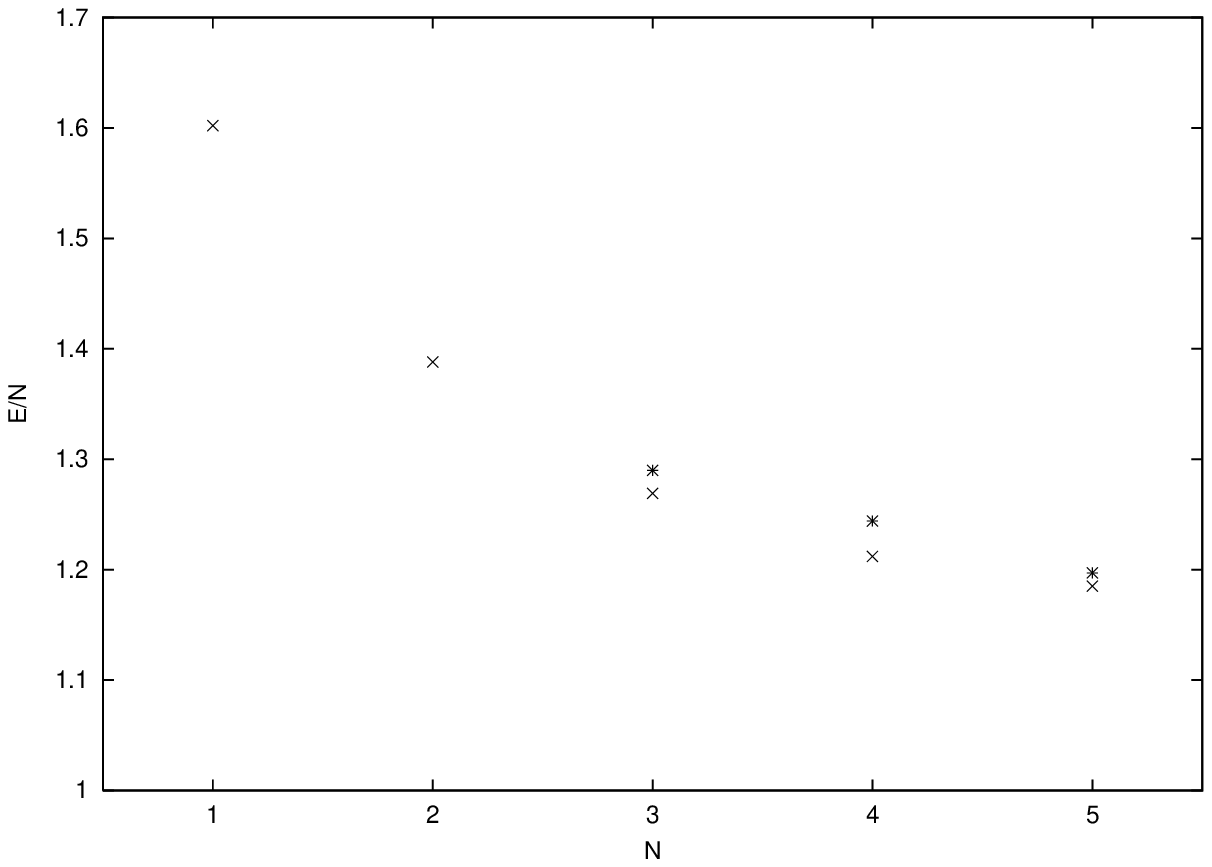}
\caption{The energy per monopole $E/N$ for the axially
symmetric monopoles (crosses) and Platonic monopoles (stars).}
\end{center}
\end{figure}

The fact that for the axially symmetric solutions 
the energy per monopole decreases as a function of
increasing monopole number (we have also checked that this trend continues
up to $N=10,$ using larger grids) makes it possible that the minimal energy
monopole is axially symmetric for all $N\ge 2.$ In order to test this
we have computed some non-axially symmetric monopoles with $N>2$ which
have the symmetries of the known minimal energy Skyrmions, since these are the
obvious non-axial contenders for minimal energy monopoles.

The minimal energy $N=3$ Skyrmion has tetrahedral symmetry $T_d$ and
the relevant rational map is the one mentioned earlier,
$R=(\sqrt{3}iz^2-1)/(z^3-\sqrt{3}iz).$ Annealing produces the tetrahedral
3-monopole displayed in fig.~2E which has an energy
$E_3^T=3.869.$ This is very slightly higher than the energy of 
the axial 3-monopole $E_3=3.807,$ and since $E_3^T-E_3=0.062$ we expect
that even though this difference is almost as large as the likely 
overall error in the computation of each individual energy,
 it is an order of magnitude greater than the errors we estimate
in the comparison between two energies. This calculation suggests
that the axial 3-monopole has less energy than the tetrahedral 3-monopole,
in contrast to Skyrmions, and hence that it is likely to be the minimal
energy 3-monopole. Of course, since the energy differences are small it
is desirable to have a more accurate calculation of both these energies
using larger grids, but this is beyond our current resources. We have
verified that the axial 3-monopole has less energy than the 
tetrahedral 3-monopole for a number of other values of the parameter $\mu$
and also performed another consistency check by computing the energy
of the additional term given the two different BPS 3-monopoles.
This will be a good approximation to the excess above the BPS bound
in the limit where $\mu$ is small, so that the fields vary little
from the BPS configurations. This result is in agreement with the
full nonlinear computation since it yields an excess energy which is
slightly less for the axial 3-monopole than for the tetrahedral 3-monopole,
though we must point out that this calculation does appear to be very
sensitive to obtaining the BPS solution to a very high accuracy.
In principle, given the correspondence between BPS monopoles and
rational maps, the additional energy contribution should provide an
interesting energy function on the space of rational maps, though it
does not seem possible to obtain any explicit information about this
energy function without first computing the monopole fields, which can
only be done numerically and is computationally expensive.

An interesting question, given that our results suggest that the
tetrahedral 3-monopole is not the minimal energy solution, is whether
this is a stable local minimum or a saddle point solution.  We are
unable to answer this question at this stage, since the algorithm
requires the Higgs field to be fixed on the boundary of the grid with
a prescribed form, and hence symmetry.  In principle, since we
have not explicitly fixed a gauge, any Higgs field which has a winding
number equal to $N$ is equivalent to any other, so it should be
possible to move between different configurations if the symmetry is
initially broken by the gauge potentials, but in practice this does
not happen since the energy differences between various configurations
are too small and the gauge potentials quickly anneal to match the
symmetry of the Higgs field.  It is this
technical difficulty which prevents us from simply finding the minimal
energy $N$-monopole by starting from an asymmetric initial condition,
which is the method used for Skyrmions but in that case the Skyrme
field is fixed on the boundary of the grid to be a constant and
contains no information about the structure and symmetry of the
Skyrmion.

The minimal energy 4-Skyrmion has octahedral symmetry $O_h$ and
is described by the rational map 
$R=(z^4+2\sqrt{3}iz^2+1)/(z^4-2\sqrt{3}iz^2+1).$
Using this map we compute the cubic 4-monopole displayed in
fig.~2F with energy $E_4^O=4.974.$ 
This is again slightly larger than the energy
of the axial 4-monopole $E_4=4.847$ and further supports our findings that
the minimal energy monopoles do not share the symmetries of the 
minimal energy Skyrmions.

The minimal energy $N=5$ Skyrmion has only the dihedral symmetry $D_{2d}$
and corresponds to a rational map of the form 
$R=(z^5+bz^3+az)/(az^4-bz^2+1)$
where $a$ and $b$ are particular real constants. 
Using the values associated with the minimal energy 5-Skyrmion
produces the monopole displayed in fig.~2G with an energy
$E_5^D=5.982$ which is larger than the axial energy $E_5=5.924.$
For charge 5 there is also another obvious minimal energy
candidate, which is an octahedrally symmetric $O_h$ monopole
associated with the above rational map in which the parameter
$b$ is zero and $a=-5.$ The annealed monopole has energy 
$E_5^O=5.987$ and is presented in fig.~2H. 
Deforming the dihedral monopole to the octahedral monopole
produces a tiny change in energy, and the difference is 
even within the numerical errors expected when comparing two energies,
so we can only conclude that the numerical results suggest that
both have higher energy than the axial 5-monopole, but which of
these two has the lower energy is not clear.

For all the charges and examples discussed above we
have performed several other computations using both larger
and smaller values for the parameter $\mu$ and found qualitatively
similar results. In all cases the axially symmetric monopoles are
always those with the lowest energy, suggesting that this is the case
for all $\mu>0.$ We have also examined the replacement of the 
fourth-order Skyrme term by a sixth-order term and found similar
results.

\section{Conclusion}

Motivated by the similarities and differences between BPS monopoles
and Skyrmions we have investigated a modification of the usual
BPS Yang-Mills-Higgs model by including a Skyrme-term formed
from the covariant derivatives of the Higgs field. 
We found that this modification indeed produces monopoles which are more like 
Skyrmions, in the sense that bound states now exist, but that the
numerical results suggest that the minimal energy monopoles for 
charges greater than two do not share the symmetries of the 
minimal energy Skyrmions, but instead appear to be axially symmetric.
The energy differences we have found are not substantial,
so further more accurate computations would be desirable,
but we have demonstrated a significant difference (with values
well { beyond} our expected numerical errors) between the behaviour
of axially symmetric monopoles in our modified model and axially
symmetric Skyrmions; in the axial monopole case the energy per soliton
is a decreasing function of soliton number and in the Skyrmion case
it is an increasing function. This property alone demonstrates that
our modified monopoles have qualitative differences with Skyrmions.

There are a number of interesting properties of monopoles in the
modified model which require further investigation. These include a
study of the energy of a 2-monopole configuration as a function of
the monopoles separation and the related issue of how the interaction
between two well-separated monopoles depends on their relative
phase. The dynamics and scattering of monopoles in this model would
also seem worth investigating, both using full field simulations and
approximate techniques. In principle the moduli space approximation could
be applied to Skyrmed monopoles by treating the modification as a
perturbation to the BPS monopole metric together with an
induced potential function on the BPS monopole moduli space.

Although the main motivation for this work is to explore connections
between various types of three-dimensional topological solitons, the
additional Skyrme term that we have included is a natural modification
that might arise in an effective theory. In this context the value
of $\mu$ is expected to be much smaller than the value we have studied
for numerical convenience, but the qualitative features of our results,
 such as monopole bound states, remain valid.

\section*{Acknowledgements}
Many thanks to R. Flume, C.J. Houghton, B. Kleihaus, J. Kunz and
 N.S. Manton for useful discussions. We thank Enterprise--Ireland and
 the British Council for financial support under project
 BC/2001/021. PMS acknowledges the EPSRC for an Advanced
 Fellowship. The research of DG is supported by 
Enterprise--Ireland grant SC/2000/020.

\end{document}